\PassOptionsToPackage{table}{xcolor}

\documentclass[acmsmall]{acmart}

\usepackage[table]{xcolor}

\definecolor{tablegray}{gray}{0.92} 

\AtBeginDocument{%
  \providecommand\BibTeX{{%
    \normalfont B\kern-0.5em{\scshape i\kern-0.25em b}\kern-0.8em\TeX}}}

\setcopyright{acmcopyright}
\acmJournal{PACMHCI}
\copyrightyear{2024}
\acmYear{2024}
\acmVolume{8} \acmNumber{CSCW} \acmArticle{01} \acmMonth{03}
\acmDOI{XXXXXXX.XXXXXXX}
\acmPrice{15.00}
\acmISBN{978-1-4503-XXXX-X/18/06}

\begin{document}

\title{Dismantling Gender Blindness in Online Discussion of a Crime/Gender Dichotomy}

\author{Yigang Qin}
\orcid{0000-0001-7843-2266}
\affiliation{%
  \institution{Syracuse University}
  \city{Syracuse}
  \country{United States}}
\email{yqin27@syr.edu}

\author{Weilun Duan}
\orcid{0000-0003-4107-2302}
\affiliation{%
  \institution{City University of Hong Kong}
  \city{Hong Kong}
  \country{China}}
\email{weiluduan2-c@my.cityu.edu.hk}

\author{Qunfang Wu}
\orcid{0000-0003-4991-1393}
\affiliation{%
  \institution{Harvard University}
  \city{Cambridge}
  \country{United States}}
\email{qunfangwu@fas.harvard.edu}

\author{Zhicong Lu}
\orcid{0000-0002-7761-6351}
\affiliation{%
  \institution{City University of Hong Kong}
  \city{Hong Kong}
  \country{China}}
\email{zhicong.lu@cityu.edu.hk}

\renewcommand{\shortauthors}{Yigang Qin et al.}

\begin{abstract}

Contemporary feminists utilize social media for activism, while backlashes come along.  The gender-related discourses are often diminished when addressing public events regarding sexism and gender inequality on social media platforms. The dichotomized debate around the Tangshan beating incident in China epitomized how criminal interpretations of gender-related violence became a backlash against feminist expressions. By analyzing posts on Weibo using mixed methods, we describe the emerging discursive patterns around crime and gender, uncovering the inherent gender-blind sexism that refutes feminist discourses on the social platform. We also highlight the critical restrictions facing grassroots feminist activism in Chinese cyberspace and propose implications for the design and research related to digital feminist activism.
\end{abstract}



\sloppy

\maketitle

\section{Introduction}

Throughout nearly two centuries, feminists across the globe have achieved remarkable milestones in their relentless pursuit of gender equality and justice. Rooted in the tradition of women's obedience to men, improving the status of women in China has much work ahead \cite{xiongya_women_existing_2003}. Although promoting women's rights in various fields is an officially sanctioned policy \cite{translate_prc_2022}, subtle forms of gender oppression were universal and rarely acknowledged by Chinese officials or public media. The chained mother of eight in Fengxian, representing the worrying status of human trafficking of women, was minimally reported in Chinese domestic media \cite{kuo_plight_2022}. The Shanghai ``Little Red Mansion'', a case of underground forced sex work, was reported exclusively about gangs and corruption, followed with little to no institutional action or official response to support women and guarantee their basic rights \cite{koetse_uncovering_2021}. In another case where a former Chinese tennis player reported the former vice premier of China pressuring her into sex, Chinese media outlets remained silent \cite{myers_chinese_2021}.

While the authority has continuously avoided commenting on women's status in those cases, the Chinese public has adapted to the new online avenue where gender issues have become a daily agenda. They are increasingly involved in debates around feminism and gender issues, concurrent with the rising popularity of feminism \cite{mao_feminist_2020}.  However, often overlooked is that gender factors are usually either neglected or downplayed in discussing public events representing women's experiences and gender inequality. Feminism incurs backlashes partially because of Chinese cultural and political norms \cite{chai_short_2022, yang_power_2009, yang_civil_2018}, which resembles the emergence of anti-feminism in Western nations in the post-feminist era \cite{lazar_feminist_2007}. In the most straightforward form, grassroots feminists are often stigmatized as those who seek privileges for women over men or ``incite conflict between genders.'' Within such a dominant environment hostile to feminism, the status of women would be hard to improve. Prior work in HCI investigated artifact and platform designs informed by feminist sensibility \cite{bardzell_feminist_2010, fiesler_archive_2016}. Building on that, in this study, we intended to understand how digital feminist activists and other online populations frame gender issues and feminist agendas in discussing public events where the gender factor is entangled with other social issues. Here, \textit{framing} is defined as the way in which people deliver ideas and worldviews through utterance~\cite{lakoff2014all}.

We chose the Tangshan incident as the case to bring diverse public opinions into the foreground. The Tangshan beating incident happened in the city of Tangshan in China, where a group of gangster men attempted to harass and resorted to fierce violence toward a woman in a midnight barbecue restaurant. After surveillance footage was uploaded online, the incident quickly incited widespread sentiments and discussions on Weibo\footnote{Weibo (stands for micro-blog in English), or Sina Weibo, is China's most popular micro-blogging platform.} about social security, public administration, sexual harassment, and gender inequality. Among the discussions, however, women's experiences and feminist topics were subject to ignorance and denial by a large portion of the public \cite{wang_brutal_2022}. To investigate the case, we collected online posts associated with the Tangshan incident from Weibo and aimed to answer the following research questions: 

\begin{itemize}
    \item \textbf{RQ1}: What were the trends and topics in discussing the Tangshan incident?
    \item \textbf{RQ2}: What discursive patterns emerged in discussing the gender dimension of the incident?
\end{itemize}

In total, we collected 724,446 posts on Weibo discussing the Tangshan incident. A mixed-methods approach was used. To answer RQ1, we analyzed the uses of hashtags in the posts, followed by developing a fine-tuned machine-learning classifier to classify all the posts into three categories: (1) \textit{Crime-related}, (2) \textit{Gender-related}, and (3) \textit{Irrelevant or Ineffective}. Next, topic modeling was run for each category to reveal popular topics. To answer RQ2, we applied feminist critical discourse analysis (FCDA) on the \textit{Crime-related} and \textit{Gender-related} posts to reveal discursive patterns.

The findings demonstrated how Chinese digital feminist expressions encountered backlashes from normative criminal and governmental framing of the incident. The discursive patterns of such framing dictated the recently-emerged ideology of \textit{gender-blind sexism} \cite{stoll2013race} that attempts to deny gender inequality. This further testified that overt and radical sexism has partially transformed into subtle forms, whose trajectory resembles how racism has shifted in the past decades \cite{augoustinos_language_2007, zamudio_traditional_2006, wu_conversations_2022}. On the other hand, Chinese grassroots feminists advocated for the validity of openly discussing gender issues online, resisted such gender blindness, but also experienced the frustration of participating in digital activism amidst cultural and political restrictions in China.

This work's key contributions to the CSCW community include (1) offering an empirical case study of the divergent online framing of a gender-related violence incident in a non-Western context and (2) proposing design and research implications based on highlighting the challenges facing digital feminist activists on Chinese social media.

All quotes this paper were translated from Chinese (unless expressly noted) with identifiers removed. \textbf{Content Warning}: Quotes may contain offensive language that may be uncomfortable to some readers.

\section{Related Work}

\subsection{Gendered Online Discourses and Collective Actions}
Social media, a crucial channel of civic participation \cite{semaan_navigating_2015, wulf_ground_2013}, has been studied extensively in previous HCI and CSCW work focusing on content and form of interaction between users and platforms. These studies inspected public discourses on social media to understand the online representation of specific social issues, such as political elections \cite{vigil-hayes_complex_2019}, migrants and refugees \cite{khatua_struggle_2021}, veterans \cite{zhou_veteran_2022}, racial protest and social movement \cite{twyman_black_2017, tufekci_social_2012, jackson_hijacking_2015, choudhury_social_2016, freelon_beyond_2016, jackson_ferguson_2016, le_compte_its_2021, starbird_how_2012, borge-holthoefer_content_2015}, identities and communities \cite{dym_coming_2019,dosono_moderation_2019}. As the product of people's interactions, online public discourses provide a valuable resource for understanding those complex social dynamics and letting us reflect on the implications of digital platforms.

This paper focuses on gender issues in public discourses. As suggested by previous work, online discourses are gendered, on the one hand, meaning that differences exist between men's and women's expressive rhetoric in online spaces; on the other hand, implying inequalities and politics behind the expressions. Compared to men's assertive discursive style, women tend to possess cooperative, supportive discourses and affiliation through interactions with others \cite{hayat_gendered_2017}. The assertive discursive style of men comes along with the pervasive toxic disciplinary rhetoric towards women \cite{cole_its_2015}. But on the other hand, platform features like hashtags can be liberating for women in that feminist hashtags and rhetoric can call into question rhetorical agency and power relation \cite{cole_its_2015}.  Women's disciplined online experience was also evident in cases demonstrating toxic masculinity \cite{massanari_gamergate_2017} and online harassment \cite{blackwell_when_2018, jhaver_online_2018}. However, what is valuable yet less explored is how public discourses around gender could tell us the gendered facets of social platforms, not to mention the contextual particularity of women's issues in non-Western regions.

Public discourses, when produced in collective action by heterogeneous groups and individuals with certain shared aspects, constitute collective action frames \cite{snow2004framing}. HCI and CSCW researchers examined how digital media sustains social movements. Information seeking, dissemination \cite{wulf_ground_2013}, framing \cite{crivellaro_pool_2014, dimond2013hollaback}, solidarity/union building \cite{starbird_how_2012, state_diffusion_2015}, and counter-narrative building \cite{al-ani_egyptian_2012} are common usages of social platforms for sustaining social movement. Meanwhile, social movements connect people through building trust and social support \cite{khairina_trust-building_2021} and often forming a collective identity \cite{flesher_fominaya_collective_2010, ackland_online_2011}. Besides mediating visible public discourses and actions at a large scale, social platforms also support undercurrents or invisible work (e.g., organizing a smaller group of participants) critical for the movement, as shown in the Umbrella Movement \cite{kow_mediating_2016}. Bennett and Segerberg further proposed the idea of \textit{connective action} where individuals are connected loosely without ties to formal organizations yet pursuing the same goals as distinguishable to traditional collective action that forms collective identity \cite{bennett_logic_2015}. Social computing researchers also examined the development of connective action regarding different social issues mediated by online social platforms \cite{vigil-hayes_indigenous_2017, Mirbabaie2021development}.

Through these ways, social media, with all of its affordances, facilitates producing, circulating, and consuming public discourses, which are essentially heterogeneous and collective, thus holding the potential for bottom-up political change. This understanding, with the characteristics of gendered online discourses, informs the intellectual direction for this work to understand the role of social platforms in mediating a collective gender debate.

\subsection{Digital Feminist Activism in China}
A new generation of Chinese feminists is increasingly utilizing social media to grapple with disciplines of feminist expression in offline settings. Feminism groups, like Feminist Voice and Women's Awakening \cite{wang_chinese_2019}, were active on social media like Weibo and became the pioneers in publicizing feminism in China. Because of easy access to the Internet infrastructures and services, digital feminist activities can also be initiated by grassroots individuals with feminist awareness besides a small number of feminist leaders or organizations \cite{mao_feminist_2020}. Those grassroots feminists played a vital role in the feminist agenda by pushing against sexism in cultural and media respect, though less in political rights.

\paragraph{\#MeToo Movement}
A typical example of online feminist collective action is the \#MeToo movement, which is closely relevant to the Tangshan incident and thus helpful to relate. The \#MeToo movement commenced in 2017 and is widely known as a movement against sexual violence and sexual harassment. Although the term "Me Too" was first used by the African American activist Tarana Burke in 2006, the movement was initiated by the actress Alyssa Milano and later emerged in other countries. Countless people on social media with different backgrounds shared their personal yet similar experiences of being sexually assaulted, expressed empathy and encouragement to the victims, and persuaded others to continue the movement \cite{manikonda_twitter_2018}. January 2018, the movement started in China and became a way of empowering Chinese women \cite{lin_individual_2019}. Disclosures and discussions on social media encouraged them to speak up, raised awareness of sexual harassment among the public, enabled self-rescue for women, reconstructed their identity, and enabled online and offline interpersonal support and actions \cite{lin_individual_2019}.

That said, digital feminist activism, not only in China, has been criticized for not accounting for the intersectional status that emphasizes multiple patterns of oppression along dimensions such as gender, race, socioeconomic status, sexuality, and disability~\cite{crenshaw_mapping_1991}. Just as the third-wave feminism's criticism toward the second-wave of being White and elitist, digital feminist activism in the West is dominated by White participants, leaving other voices less heard and their particular issues unsolved~\cite{daniels2015trouble}. This problem remains in the Global South. Nanditha provided an empirical analysis of how the \#MeToo movement in India remained "exclusive, divisive, and fragmented" as it failed to capture and speak for the experience of Dalit and lower caste women, trans women, the LGBT community, and rural communities but instead was dominated by urban and elite Indian women's discourses~\cite{nanditha_exclusion_2022}.
Studies of the Chinese \#MeToo movement also reveal the biased demographics in that feminist and pro-change liberal participants were urban, well-educated, and middle-class groups~\cite{Zeng2019, yin_intersectional_2021}. Rural and working-class people, though constituting a significant portion of the Chinese population, were underrepresented in the movement, but unfortunately more vulnerable to sexual assaults and get unreported~\cite{yin_intersectional_2021}. The movement was, therefore, neither inclusive nor had harnessed the potential of mobilizing the subaltern population for social change~\cite{yin_intersectional_2021}.

\paragraph{Backlash} 
Because of the sensitive domestic policy about social movement in China \cite{chai_short_2022, yang_power_2009} and political discourses for Internet governance (e.g., \textit{wenming}, meaning civility) \cite{yang_civil_2018}, Chinese feminists face backlash and suppression from the authority and part of the public. Yin and Sun identified two sources of backlash during the Chinese \#MeToo movement: national censorship and misogyny \cite{yin_intersectional_2021}. The government or the platform conducts censorship to avoid large-scale collective action. It filters out sensitive words and blocks associated accounts. Social media users can only modify their posts to circumvent censorship \cite{zeng_metoo_2020, zhang_sensitive_2019}. The other backlash is from anti-feminist, misogynistic, or male chauvinist groups. Similar situations happened in other nations in the post-feminist era \cite{puente_twitter_2021, banet2016MasculinitySoFragile}. Online feminist activism may even strengthen these thoughts, intensifying negative impressions towards feminists \cite{han_searching_2018}. The backlash also parallels the stigmatization, usually claiming that Chinese feminists only seek privileges over men but do not perform duties \cite{mao_feminist_2020}.

\subsection{Gender-Blind Framing in Online Public Discourse}
Gender blindness, or gender neutrality, is an ideology that discounts gender as a salient factor in how people experience inequalities daily. It is a notion derived from liberal feminism, which downplays gender differences~\cite{stoll2013race, essed_companion_2009}. Similar to gender blindness, color blindness is used to conceal racist ideologies and maintain the status quo that privileges White people. Bonilla-Silva argued that color blindness is the primary mechanism to frame subtle and covert racism in the contemporary U.S. Bonilla-Silva articulated four color-blind racial frames~\cite{bonilla2018racism}. They are \textit{abstract liberalism}, \textit{naturalization}, \textit{cultural racism}, and \textit{minimization of racism}, which generally argue that race is not the significant factor for social inequalities related to race. For example, the \textit{minimization of racism} frame minimizes the impacts of structural racism on the living conditions of People of Color. If People of Color challenge White privilege, they might be called ``playing the race card''~\cite{bonilla2018racism}. White people utilize the four color-blind racial frames to deny their racist ideologies at all levels of society. The ideologies of gender-blind sexism can be seen as another online disciplinary rhetoric - asserting abstract liberal and cultural explanations of social differences between genders, minimizing, and naturalizing those differences \cite{anskat_dissemination_2021}.

Referring to gender-blind sexist frames, we investigate how gender-related frames were developed online during the Tangshan incident and their relations to gender-blind sexism. Prior studies have explored gender-blind framing in various contexts, such as rape incidents~\cite{stoll2021effects}, academia~\cite{shukla2022persistence}, and menopause~\cite{lazar_parting_2019}. This work explores how gender-blind framing was constructed on a social media platform (i.e., Weibo) and how grassroots feminists generated counter-frames to challenge this framing. As Rode contended, technology is gendered, reflecting a dominant masculine culture~\cite{rode2011theoretical}. The social media platform's culture shapes the public discourses and is also shaped by the public discourses. This work further examines the relationship by analyzing different topics and discursive patterns around the gender dimension.

\section{Background: A Gender-Related Violence Incident}

On June 10, 2022, a group of men assaulted four women in a barbecue restaurant in Tangshan, China. Before dawn, a man attempted to sexually harass a woman. When faced with resistance, he and his accomplices assaulted the woman and her companions in front of others in the restaurant \cite{a-tangshantoutiao}. 
The video footage from the scene gained widespread attention after being posted on Weibo. People closely followed the case's progress and called for strictly punishing the perpetrators, paying attention to the victims' injuries, and enhancing social security. As more information about the case emerged, such as the assailants being members of a local gang, online discussions shifted towards gangs and their possible connections within the government. Following the incident, the Tangshan police announced the implementation of ``Operation Thunder,'' an initiative aimed at eliminating local gangs \cite{a-thunder}.

Simultaneously, discussions on gender issues stirred up controversy on the internet. Some focused on gender-based violence exemplified by the incident, while those concentrating on gangs and corruption criticized those who linked the incident to gender, creating a hostile online atmosphere \cite{a-genderdiscussion}. Weibo blocked over 200 accounts under the pretext of ``inciting conflict between genders'' \cite{a-weibo}.

\section{Data and Methods}

We collected the data by web-scraping publicly available posts from Weibo. Weibo became the online hub for communicating user-generated and official media content in mainland China after its release in 2009 \cite{shanshan_userbehvr_2021}. Weibo substituted Twitter as a permitted domestic microblog site because Twitter was officially blocked from access by China's authority in 2009 \cite{bamman_censorship_2012}.

\subsection{Data Collection and Processing}
We first collected 988 unique hashtags containing the keyword ``Tangshan'' from Weibo's full-site search engine to prepare for web scraping. We chose this keyword for its frequent appearance in discussing the event based on the best of our knowledge. Then, the first author manually filtered out irrelevant hashtags to the event, which the second author reviewed. They discussed different decisions and agreed on a final list of 100 hashtags and corresponding selection criteria. The hashtags constitute the following categories.

\begin{itemize}
  \item Reports of incident detail and progress
  \item Public comments on the incident
  \item Response from media, government, organizations, celebrities, witnesses, and others
  \item Latest status of the victims
  \item Suspects' personal information
\end{itemize}

We scraped posts containing these shortlisted hashtags from Weibo. The web crawler was developed using the Scrapy framework \footnote{Website of Scrapy: https://scrapy.org/}. Posts and reposts mentioning any hashtag in the list were collected. The top five hashtags in the list (sorted by the number of discussions displayed on Weibo) are shown in \autoref{top5htag}. In total, 724,446 individual posts and reposts with any of the hashtags were collected and stored in a local database. The collected information was public and accessible to any Weibo users. The information scraped contains (1) message ID, (2) user ID, (3) post/repost time, (4) post/repost IP location (disclosing province-level IP location on social media is mandatory in China \cite{dong_chinas_2022}), (5) user's gender, (6) text content, and (7) those of the reposted post (parent post). The time scope of the data spans from the first day of the report---12 AM, June 10, 2022, to three weeks later at 11:59:59 PM, June 30, 2022. We then processed the raw text content; HTML tags, hashtags, links, emojis, special characters, and the appended parent posts were removed; punctuation, numbers, and English words were reserved. After processing, only the original text content was kept. In the end, 78,582 (10.85\%) posts became empty, and the remaining 645,864 (89.15\%) posts were used in the following analysis.

\begin{table}[htbp]
\centering
\rowcolors{2}{white}{tablegray}
\begin{tabular}{p{0.4\textwidth}p{0.2\textwidth}p{0.2\textwidth}}
\toprule
Hashtag (Translated) & Discussions & Readings \\
\midrule
\#\textit{Tangshan Beating} & 4,060 K & 4,600 M\\
\#\textit{Men beat up women in a barbecue restaurant in Tangshan} & 1,241 K & 2,240 M \\
\#\textit{Public security bureau is arresting suspects in the Tangshan beating incident} & 1,115 K & 1,770 M \\
\#\textit{Violent attack of Tangshan girls is a nightmare for all} & 704 K & 1,950 M \\
\#\textit{Tangshan beating incident all 9 people arrested} & 682 K	& 2,170 M \\
\bottomrule
\end{tabular}
\caption{Top five hashtags related to the Tangshan incident. The statistics were collected from Weibo and the last date of collection is June 25, 2022. K=1,000; M=1,000,000.}
\label{top5htag}
\end{table}

\subsection{Content Classification and Topic Modeling (RQ1)}
Two major groups of content emerged from the discussion of the incident on Weibo: (1) crime-related content and (2) gender-related content. They are distinct from each other in that the posts interpreted the event in different directions. We needed a method to classify the posts to delineate this content divide. Instead of manually classifying a sample from our database, we leveraged machine learning to classify posts at scale automatically. There are two reasons behind our employment of machine learning for classification. First, the manual classification of a sample was less favored than comprehensively classifying the mass online discussion. By using machine learning, we could quickly and feasibly obtain information about the scale and temporal trends, necessary to capture the landscape of online discourses. Second, classifying all the posts instead of a sample could facilitate our subsequent qualitative analysis for RQ2 for the following reason. When manually classifying the posts, we found a significant portion of posts irrelevant to the event or lacking substantial content. If we had sampled more posts for qualitative analysis, much time and effort would have been spent identifying and filtering these posts, and less time would have been for analyzing posts of our interest. Out of such concern for efficiency, we trained a machine learning classifier first to identify and filter out the content to facilitate our qualitative analysis and improve its comprehensiveness.

\paragraph{Data Annotation} We manually labeled a set of posts as the training set, which was used to train a natural language processing (NLP) model to classify the remaining posts automatically. The labeling process consisted of two phases. In the first phase, we randomly sampled 1,000 posts from all the non-empty posts. Two researchers independently coded the posts in an open-coding fashion to generate initial codes. Then, they discussed the codes and generated a preliminary labeling scheme that groups the codes to represent different content classes. Next, the two researchers each deductively labeled the coded posts according to the preliminary labeling scheme. This step was iterative; they revised the labeling scheme in each iteration. The first labeling's inter-rater Cohen's kappa score was 0.55, the second labeling's score was 0.69, and the third time was 0.93; until then, they stopped iteration. The final labeling scheme inductively defines three content classes as described in \autoref{codebook}. Posts with both major classes of content (\textit{Crime-related} and \textit{Gender-related}) were labeled as one of them according to the posts' emphasis on one class over the other. It was possible that we inappropriately labeled some posts with comparable inclinations to both classes, but the number of such posts was small, and regardless of which classes they went to, they would be examined in greater detail in answering RQ2. In the second phase, the second author additionally randomly sampled 3,800 posts (not overlapping with the first sample; the number was inferred from the class distribution of the first sample) from the non-empty posts and labeled them according to the final labeling scheme; the labeling was reviewed and discussed until agreement. There was an imbalance across the three classes, so down-sampling was applied to the \textit{Irrelevant or Ineffective} and \textit{Crime-related} classes to balance the dataset for training the classifier. Ultimately, 2,100 labeled posts with 700 posts in each class were used for training (80\%) and testing (20\%) the classifier.

\begin{table}[htbp]
\centering
\rowcolors{2}{white}{tablegray}
\begin{tabular}{p{0.2\textwidth} p{0.6\textwidth}}
\toprule
Content Class & Description \\
\midrule
Crime-related & Focusing on legal, criminal, and social security issues around the event, including mainstream media reports.\\

Gender-related & Talking about gender differentials, gender inequality, women's rights, sexual harassment, and gender-based violence.\\

Irrelevant or Ineffective & Asking for the latest states and progress without discussion, seeking exposure, venting emotions, spreading rumors, asking for refuting the rumors, having low-quality, describing personal conjectures, and questioning the event details.\\
\bottomrule
\end{tabular}
\caption{Labeling scheme for classification}
\label{codebook}
\end{table}

\paragraph{Content Classification} We used a fine-tuned BERT (Bidirectional Encoder Representations from Transformers) model \cite{devlin_bert_2019} to classify the remaining posts in our database. The BERT model is a state-of-the-art machine learning model for NLP and was widely applied in various research scenarios. Apart from directly using the pre-trained model, users can further pre-train the model on the task corpus and fine-tune it in a downstream task. This way of training usually yields higher performance \cite{sun_how_2020}. The model we used was based on the pre-trained BERT-WWM (Whole Word Masking) model developed specifically for Chinese text \cite{cui_pre-training_2021}. We further pre-trained the base BERT-WWM model using all the non-empty posts on a Masked-Language Modeling (MLM) task, the primary method used to train the original BERT model \cite{devlin_bert_2019}. The further pre-training step, called within-task pre-training, was useful in our case as it adjusted the BERT's network weights to fit our particular corpus. The further pre-trained model was then fine-tuned on a text classification task using the text and labels of 1,680 posts (80\% of the 2,100 labeled posts). Common classification metrics were calculated to evaluate the model's performance. The model achieved an average accuracy of 0.929 and a macro F1 score of 0.901 on the test set (20\% of labeled posts). Finally, the fine-tuned BERT model predicted the labels of the remaining non-empty posts.

\paragraph{Evaluation of Hashtag Uses} We then explored the prediction results to have an overall sense and noticed a mismatch between the hashtags people use and the actual text content posted. To explore the possible heterogeneity between the \textit{Crime-related} and \textit{Gender-related} posts' use of popular hashtags, we attempted to quantify the semantic similarity of posts' text content and the hashtags used. Another Sentence-BERT model fine-tuned on all the non-empty posts based on a pre-trained multilingual SentenceTransformers model (we also followed their instructions of fine-tuning) \cite{reimers-2020-multilingual-sentence-bert} was used to generate sentence embeddings for both the hashtags and the text content (viewing each hashtag as a sentence). Cosine similarities between the embeddings were calculated to measure semantic differences between the hashtags and text content. Scores for posts with multiple hashtags were calculated by uniformly averaging the similarity scores of each combination between the hashtags and the text. To reduce randomness, we trained five such models and averaged their produced similarity scores.

\paragraph{Topic Modeling} Topics of the \textit{Crime-related}, \textit{Gender-related}, and \textit{Irrelevant or Ineffective} classes were extracted using the Gibbs Sampling algorithm for the Dirichlet Multinomial Mixture model (GSDMM) \cite{yin_dirichlet_2014}. GSDMM performs better than Latent Dirichlet Allocation (LDA) for modeling short text like social media posts because GSDMM assumes one topic for one document, which is usually true for micro-blogs. We tested different choices of number of iterations (\textit{n}), number of topics (\textit{k}), $\alpha$ and $\beta$ value (hyper-parameters for the model). According to the original paper, the number of clusters converged at \textit{n}=30 with good performance across diverse testing datasets, which can be transferred to analyzing our dataset;  $\alpha$=.1 also produced good results across the testing datasets; $\beta$=.1 produced balanced performances and relatively fewer clusters. \textit{k} values were set initially a large enough initial value (\textit{k}=50) and the model converged at around \textit{k}=15 for the \textit{Crime-related} and \textit{Gender-related} class and \textit{k}=20 for \textit{Irrelevant or Ineffective} class, but the topics were difficult to interpret, so we gradually reduced the \textit{k} values until the clusters remained cohesive within, separate between, and easy to interpret and present \cite{yin_dirichlet_2014}. We finally chose \textit{n}=30, \textit{k}=4, $\alpha$=.1, $\beta$=.1, for the \textit{Crime-related} and \textit{Gender-related} classes; \textit{n}=30, \textit{k}=10, $\alpha$ =.1, $\beta$=.1, for the \textit{Irrelevant or Ineffective} class. We assigned each topic a name based on the keywords and text content.

\subsection{Critical Discourse Analysis (RQ2)}
Our qualitative method incorporated critical discourse analysis with feminist social theories similar to \cite{lazar_parting_2019} to extract and assess emerging discursive patterns in the posts. Based on our previous content classification, we started with an inductive thematic analysis \cite{braun_using_2006} for posts in \textit{Crime-related} and \textit{Gender-related} classes. Since women's experiences, rights, and gender-related topics were at the center of discussion, we adopted the methodology of feminist social science \cite{bardzell_towards_2011}, and in operation, primarily informed by and followed the principles of feminist critical discourse analysis (FCDA) proposed by Michelle Lazar \cite{lazar_feminist_2007}. To appropriately refer to people's gender, we use gender identities (i.e., man and woman) to substitute more biological-oriented terms (i.e., male and female).

The corpus for thematic analysis consisted of first, 1,400 out of 2,100 posts used in training the classifier (700 posts of the \textit{Crime-related} class and 700 of the \textit{Gender-related} class), and second, 1,000 additional posts randomly sampled from the two classes (500 each), resulting in a total of 2,400 posts. We sampled from the predicted classes of posts because, as mentioned previously, this made our analysis efficient and effective. The two researchers who labeled these posts performed the thematic analysis in an open-coding fashion but periodically discussed new findings and questions with other authors. After constructing themes from the posts, we further randomly sampled 50 posts (25 each) from the two classes to examine if we had already achieved thematic saturation where no new themes emerged from the data \cite{hagaman_howmany_2017}. We ended coding after the first test of saturation. Example codes included \textit{personal experience} - usually experience of being sexually harassed, \textit{feminist advocacy} - call for improving women's rights, \textit{denouncement} - criticism to the abusers, \textit{analysis} - discussion about aspects of the incident, and \textit{incident details} - telling or making conjectures about the incident details. 

We then critically assessed the themes with a focus on the ways of framing with high sensitivity, constantly drawing connections to other events, concepts, and the sociopolitical dynamics behind. It thus allowed us to depict the multiple nuanced facets of the online discourses from a critical lens. The results of this discourse analysis were iteratively reviewed and revised by all the authors to avoid inappropriate or biased interpretations.

We especially analyzed the discourses from the perspective of feminism, especially the third wave \cite{lazar_feminist_2007}. It is important to point out that we do not use the term ``feminism'' simply in the sense of general political advocacy for women's socio-cultural status, but rather more towards a dialectical social research practice that integrates feminist epistemology, methodology, and methods. As elaborated in \cite{bardzell_towards_2011}, a feminist epistemology allows us to be reflexive on the positions (in terms of marginal and central) and views of researchers as knowledge producers and how these affect the scientific objectivity and moral objectivity of our science. A feminist methodology is an implementation of the epistemology, is inherent to the feminist values, positions, and perspectives, and informs the selection and conduct of research methods. Lastly, feminist methods should not be seen as standalone or special methods that are naturally feminist but rather as normative methods used by researchers who adhere to feminist epistemology and methodology in their research. Here the critical discourse analysis method was used under the motivation of feminist methodology. Such a critical feminist standpoint deepened and enriched our insights, illuminating a broader background of women's issues as manifested in the discourses. Although we do not deny that this stand could have limited our intellectual reach, feminist scrutiny was pivotal to this incident and the broader social issues it embodies.

\subsection{Positionality Statement}
The data analysis involved iterative discussions to ensure that the researchers' positionality would not bias the data analysis. We must acknowledge that the researchers' experiences and opinions cannot represent everyone's. We reflect on our position and privilege as scholars.

All four authors are Chinese who are interested in and practice feminist HCI research. The first author, who led the analysis, has been paying attention to gender issues, especially in China, for years. He reflects on himself as privileged by male identity, but might have missed nuances when analyzing the data. Both the first and the second authors are Chinese men living mostly in China. The third author is a woman who grew up in China and has experienced gender inequalities mediated by gender-blindness in personal life and institutions. She brought her experiences into data analysis. However, as a highly educated woman, she has the privilege that might bias the data analysis. The fourth author, a Chinese man, has been conducting research on marginalized groups (e.g., rural and female users, LGBTQ+) in China for years, and he oversees the analysis based on his prior research experience.

\section{Findings}
We found that gender-related discussion around the Tangshan incident was subject to downplay and ignorance on the social platform Weibo---the "legitimate" discussion concerned crime, social security, and governance. Gender-related and feminist discourses had proponents but still were not recognized enough by the general public and even encountered negation regarding their validity. The counter-framing from Chinese grassroots feminists, on the other hand, resisted the dominant discourses through various discursive patterns.

\begin{figure}[htbp]
  \centering
  \includegraphics[width=\textwidth]{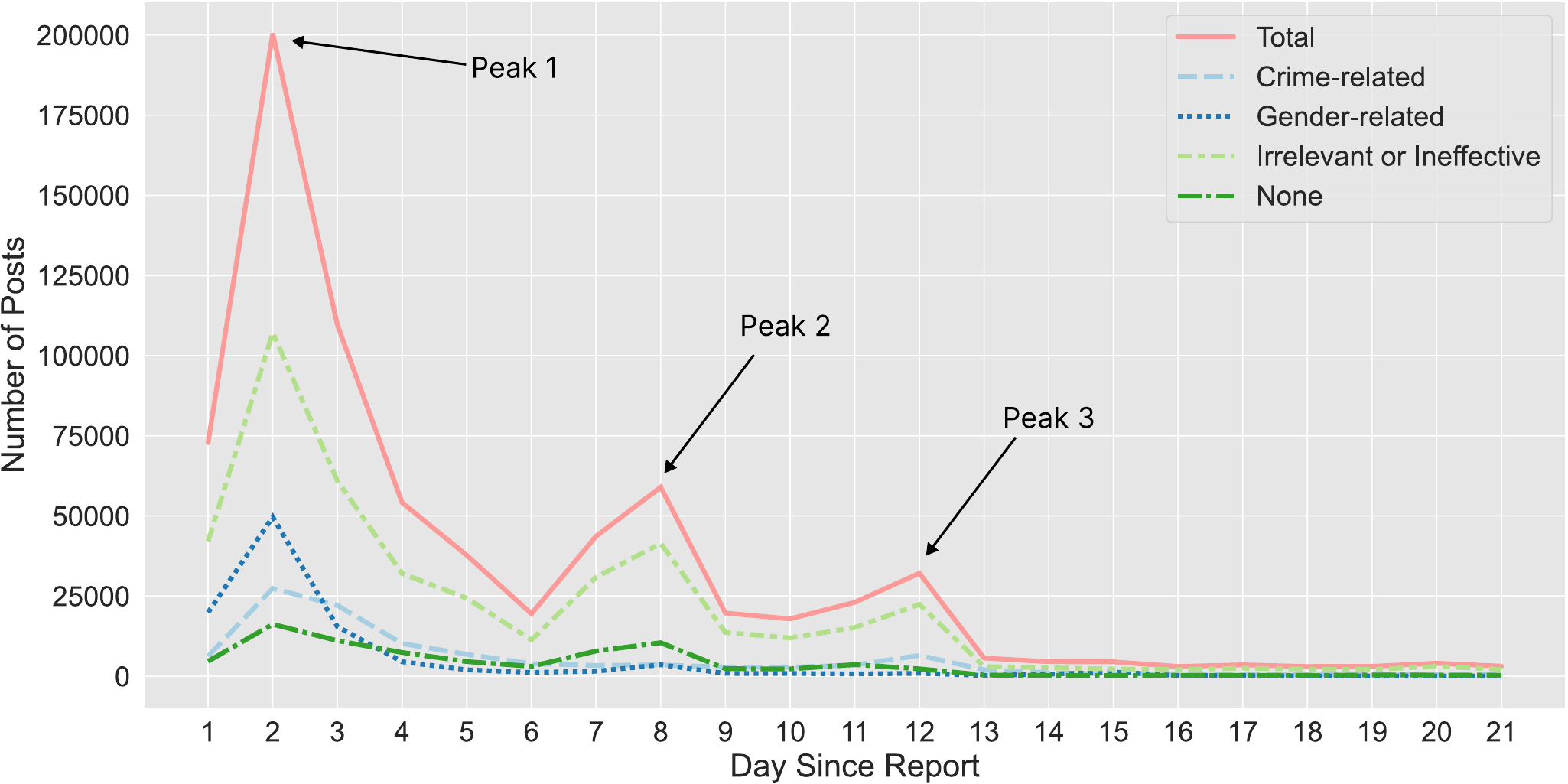}
  \caption{Temporal trend of post volume}
  \label{timeline}
\end{figure}

\subsection{RQ1: Trends and Topics of Public Discussion}
In total, the 645,864 non-empty posts were classified by the BERT model into three classes as mentioned earlier: (1) \textit{Crime-related} (106,187 posts, 16.44\%), (2) \textit{Gender-related} (104,209 posts, 16.13\%), and (3) \textit{Irrelevant or Ineffective} (435,468 posts, 67.42\%). The empty posts without original text after pre-processing were labeled as \textit{None} (78,582 posts, 10.85\%).

\autoref{timeline} shows the temporal trends of the number of posts of each class starting from the day of reporting the event until three weeks later, between June 10th and June 30th. According to the temporal statistics, discussion around this event on Weibo increased rapidly initially, then declined to negligible in two weeks. The incident attracted massive attention on the second day after the report, and the discussion reached its first peak (Peak 1) on the same day. Another two peaks (Peak 2 and 3) also happened in the first two weeks. The first peak might result from the primary discussion, while the second peak (on June 17th) could be attributed to the Tangshan government narrowing the restrictions for entering Tangshan. The third peak (on June 21st) coincided with the Tangshan government announcing that several officials undergo disciplinary investigation.

Public attention shifted during the early stage. On the first and second days after the report, gender-related topics were much more popular than crime-related ones; the situation began to reverse on the third day. The popularity of gender-related topics dropped significantly on the third day and thereafter remained nearly silent. Crime-related topics prevailed throughout the following days, including on the days of Peak 2 and 3. The temporal trends manifest the dynamics of public attention and imply the possible shift in user groups---users with gender awareness might have initially promoted the discussion, but the focus shifted to criminal issues when the broader population got involved.

\begin{figure}[htbp]
\centering
\includegraphics[width=\textwidth]{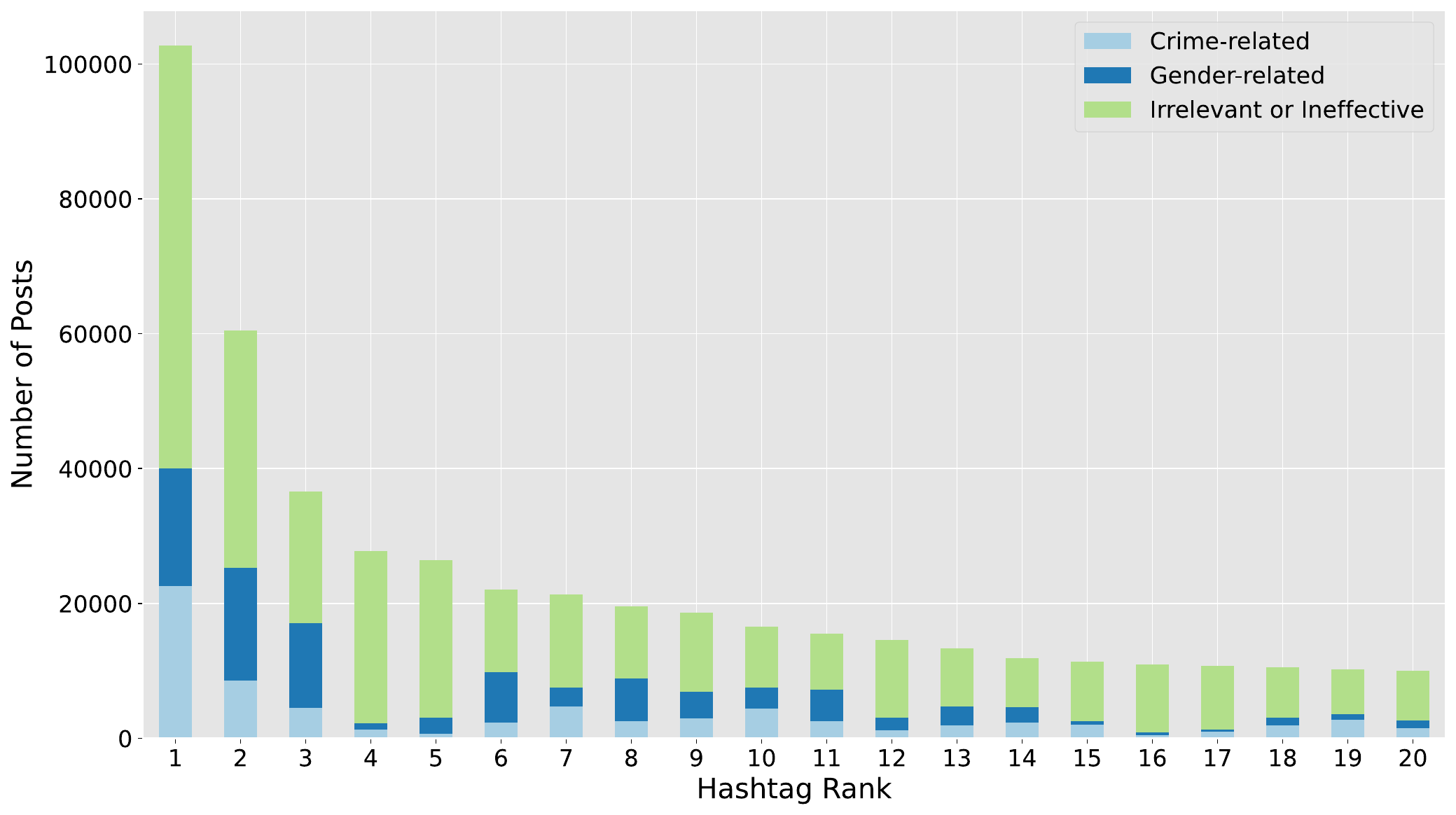}
\caption{Distribution of Posts with the Top 20 Hashtags Per Content Class}
\label{Top10Count}
\end{figure}

\begin{figure}[htbp]
\centering
\includegraphics[width=\textwidth]{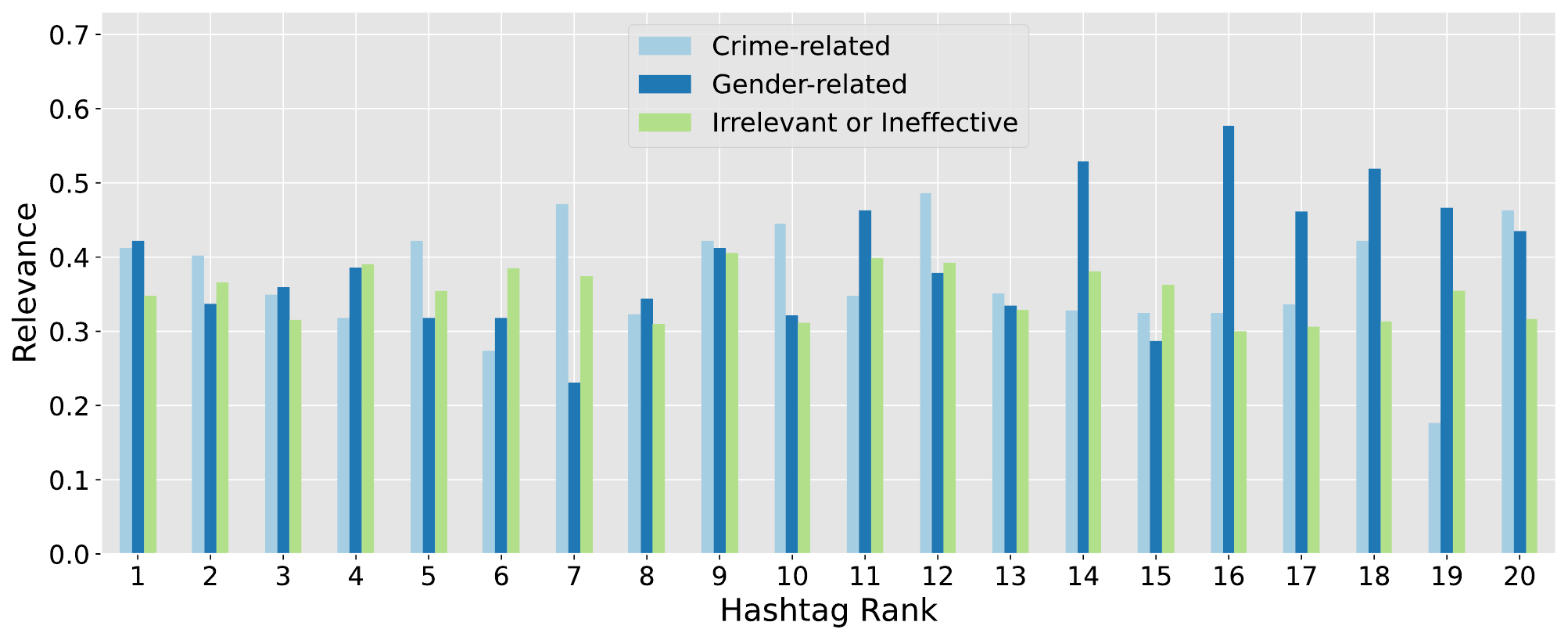}
\caption{Average Semantic Relevance Between Hashtags and Text Content of Posts with the Top 20 Hashtags Per Content Class}
\label{Top20Valence}
\end{figure}

\subsubsection{Hashtag-Text Mismatch}
The quantities of \textit{Crime-related} posts and \textit{Gender-related} posts with the top-20 hashtags were comparable in general, still posts of the \textit{Irrelevant or Ineffective} class constituted the largest portion, as illustrated in \autoref{Top10Count}. \textit{Crime-related} posts more frequently appeared, comparing to \textit{Gender-related} posts, with the 1st (\textit{\#Tangshan beating}), 7th (\textit{\#Tangshan beating incident all 9 people arrested}), and 10th (\textit{\#The Tangshan beating incident}) most popular hashtags. On the contrary, clearly more \textit{Gender-related} posts were with the 2nd (\textit{\#Men beat up women in a barbecue restaurant in Tangshan}), 3rd (\textit{\#Violent attack of Tangshan girls is a nightmare for all}), 6th (\textit{\#Several men beat up girls in a barbecue restaurant in Tangshan}), 8th (\textit{\#Severely punish the lunatics in the Tangshan barbecue restaurant beating incident}), and 11th (\textit{\#The Tangshan beating incident is a naked provocation to a society ruled by law}) mostly used hashtags. It is noticeable that many \textit{Gender-related} posts did not include hashtags with explicit inclinations toward feminism but mostly about facts or sentiments (e.g., \textit{\#Violent attack of Tangshan girls is a nightmare for all}), even though holding such opinions in their text content.

The semantic relevance scores between hashtags and text content further testified to this phenomenon, as shown in \autoref{Top20Valence}. The relevance scores indicated that \textit{Crime-related} posts semantically aligned with popular hashtags better than \textit{Gender-related} and the \textit{Irrelevant or Ineffective} posts did. \textit{Crime-related} and \textit{Gender-related} posts had comparable semantic relevance with the top 1 to 3 hashtags. Many hashtags between the 14th and 20th ranks semantically aligned better with \textit{Gender-related} posts. The weighted-mean hashtag-text relevance of \textit{Crime-related} posts (M=0.40) was significantly larger than that of \textit{Gender-related} posts (M=0.35) with the top 10 hashtags (\(z=188, p<.0001 \), also with a Cohen's D effect size of 1.05, implying its considerable practical significance (>.8). For the top 20 hashtags, the weighted-mean relevance of \textit{Crime-related} posts (M=0.39) was also significantly larger than that of \textit{Gender-related} posts (M=0.36), \(z=64, p<.0001 \), effect size of .32 (small practical significance possibly due to the involvement of more posts with average relevance). This indicated that the most popular hashtags, to some extent, were semantically divergent from \textit{Gender-related} text, which we arguably regarded as manifesting the posters' authentic intentions. The preliminary quantitative evaluation implied that \textit{Gender-related} expressions were somehow constrained on Weibo. In the following paragraphs, we turn to report the topics of discussion around this incident.

\subsubsection{Crime-related Content} Four topics covered the majority of content in the \textit{Crime-related} class, as listed in \autoref{crime topics}. Posts in the \textit{Crime-related} class regarded the incident as a criminal event that has little to do with gender issues. Over half of the posts talked about \textit{Gangs and Umbrella}. They were concerned with the social problems of gangland and associated umbrella personnel in the local government who concealed facts about gangs and colluded with them. The posters assumed the existence of organized local gangs and questioned the ability of local government agencies. They thought the government and public security bureau should strive to eliminate gangs. The issues of gangs subsequently raised conjectures about corruption under the assumption that only insiders could keep the gangsters from law enforcement and preserve them under the strict security policy in China. Posts under this topic usually expressed anxiety or determination to remove illegal gangs by strict law enforcement; the remaining posts drew more attention to the criminal facts about this incident or other cases. The second most popular topic was \textit{Situation and Progress Report}, under which media and grassroots users reported and discussed the victims' current situations and the progress of the investigation. The topic of \textit{Criminal Details and Legal Sentencing} included posts that reported or made conjectures regarding the detailed facts of the incident or judged its criminal nature according to law. Under the topic of \textit{Related Cases}, the posts described prior and similar criminal cases in relation to the suspects or government bureau, discussing the suspects' criminal records or the government's corruption and irresponsibility. 

\begin{table}[htbp]
\centering
\rowcolors{2}{white}{tablegray}
\begin{tabular}{p{0.35\textwidth}  p{0.6\textwidth} }
\toprule
Topics \& Top Keywords & Example Posts \\ \midrule

\textbf{Gangs and Umbrella (61.12\%)} 

Tangshan, incident, beat, society, power, wipe out gangs, umbrella, evil, report, severally punish, hope, law, people, eliminate evil, violence, matter, nation, gangland, real name, crime 
& 1. \textit{This is not a matter of gender confrontation or quality, but a resolute struggle between our society as a whole and the forces of darkness and evil...}

2. \textit{What protective umbrella, what black and evil forces!...}
\\

\textbf{Situation and Progress Report (18.28\%)} 

Tangshan, incident, beat, woman, barbecue, society, personnel, crime, June, case, beat up, video, situation, law, violence, the city of Tangshan, happen, police, journalist, suspect 
& 1. \textit{According to...
, a case of provocation and violent assault occurred in a barbecue restaurant... 
, which was investigated and handled by...
Public Security Bureau.}

2. \textit{Tangshan women were beaten by multiple people...
and now all nine suspects have been arrested. Subsequently, Tangshan set off a wave of real-name reports.}
\\

\textbf{Criminal Details and Legal Sentencing (13.00\%)} 

Tangshan, barbecue, beat, police, personnel, case, involved, beat up, June, suspect, woman, incident, arrested, crime, violence, the city of Tangshan, circulate, public security bureau, 10th day, branch 
& 1. \textit{... dragged the woman to the monitoring area outside the blind area. Whether there was a continued beating or indecent behavior, we do not know, what we know is that the woman laid on the ground but the other party was still not relieved.}

2. \textit{Drunkenness is not a reason for exemption... 
His behavior was suspected of constituting provocation and nuisance crime, intentional injury crime. In addition, the woman's resistance was a legitimate defense, not criminally responsible, and would not become the perpetrator of violence...}
 \\

\textbf{Related Cases (7.60\%)} 

Tangshan, several punish, incident, society, conference, beat, court, work, barbecue, people, June, wipe out gangs, according to law, assault case, hurry up, consume, citizen 
& 1. \textit{That Tangshan assailant Chen Jizhi is a local aquaculture boss, several times became an old rogue ... }

2. \textit{But the *** (name of a government official), which I have been reporting for years, has not only stayed put, but also was named ``excellent cadres'' in the ``national political and legal rectification.''}
\\

\bottomrule
\end{tabular}
\caption{Categorization of topics with examples for posts in the Crime-related class}
\label{crime topics}
\end{table}

\begin{table}[htbp]
\centering
\rowcolors{2}{white}{tablegray}
\begin{tabular}{p{0.35\textwidth}  p{0.6\textwidth} }
\toprule
Topics \& Top Keywords & Example Posts \\
\midrule

\textbf{Rights to Act Freely and Pursue Beauty (32.35\%)} 

girl, female, protect, girl, speak out, girl, go out, barbecue, hope, society, severally punish, harassment, dressed, friends, fault, scantily clad, night, matter, resist, victim 
& 1. \textit{You say women better not wear skirts in midnight, they were wearing white T's and pants, you say girls should not go out alone in midnight, they were in a group, you say girls should not go to shady places, they were in a well-lit street! They had not been wrong since the beginning!}

2. \textit{After things came out my sister told me not to wear makeup to go out at night in the future, anyone is teaching girls how to protect themselves, so why can't some men with cheap hands control themselves and respect girls... 
Makeup, dress, time, and place are not excuses for sexual harassment!}\\

\textbf{Anger, Anxiety, and Disappointment of Violence Against Women (29.57\%)} 

female, society, male, girl, incident, protect, matter, gender, Tangshan, hope, girl, man, speak out, violence, harass, happen, girl, beat, male, victim 
& 1. \textit{What I am angry about is the naked harassment of women in broad daylight... 
the unprovoked beating with physical advantage... 
the indifference of bystanders, the silence of men. Why can't we use the words male and female, why do we have to explain that this is not feminism?}

2. \textit{I lost sleep...
and woke up... 
with a muddled mind because of this confusing and horrifyingly vicious event. The impact of this event ...
is that it can be very vicarious, because they were publicly violated in the most ordinary and common state of life like each of us.}\\

\textbf{Denouncement Towards the Perpetrators and Men (27.41\%)} 

girl, female, male, man, Tangshan, beat, woman, harassment, incident, girl, female, sexual harassment, speak out, protect, society, girls, video, hope, severally punish, strike up a conversation 
& 1. \textit{Men are generally more egregious than women, see the video there was a passerby woman who several times rushed out but was stopped by her boyfriend. 
In the gang 
... it was also the white dressed woman to persuade the fight.}

2. \textit{So stop saying what ``girls should learn to protect themselves'' bullshit...
It is men who are worse than animals, including but not limited to domestic violence, rape, provocation.}
\\

\textbf{Anecdotes, Similar Cases, and Warnings (10.67\%)} 

girl, Tangshan, beat, speak out, incident, sexual harassment, your turn, f**k you, death penalty, suggest, tonight, female, male, Shanghai, Xi'an, subway, Fengxian, iron chain, opportunity 
& 1. \textit{...
A few girls playing in their own homes at night, there was a strange drunken man (a strong student athletics) who knocked on the door to play hooliganism and then began to beat people. After reporting the girls instead was lectured.}

2. \textit{...
When the pursuit was not accomplished he created sexual rumors to insult the girls' innocence...
what is the difference between this and the Tangshan beating incident?} \\

\bottomrule
\end{tabular}
\caption{Categorization of topics with examples for posts in the Gender-related class}
\label{gender topics}
\end{table}

\subsubsection{Gender-related Content} Analyzing posts in the \textit{Gender-related} class revealed four topics, as shown in \autoref{gender topics}. The first one concerned women's \textit{Rights to Act Freely and Pursue Beauty} without limitations from gender stereotypes, prejudice, or sexism. Many people were aware of the multitude of constraints disciplining women's behavior, including dressing. Chinese women may be criticized as ``\textit{scantily clad},'' a typical example of slut-shaming. In the case of Tangshan, the girl victims were accused for the same reason. Some posts attributed the sexual harassment and the following violence to the girls' dressing and attending the barbecue restaurant at midnight, which was conceived as ``inappropriate'' in the dominant cultural codes. Those opinions were criticized by users who promote women's right to act freely and avoid being treated unequally. Also, some pointed out that women's self-protection alone can not solve the problem unless men learn to respect and protect women. In addition, the phrase ``\textit{girls help girls}'' (appeared in English, not Chinese) and the idea of ``\textit{only girls protect girls}'' frequently appeared in this topic and sometimes others as well.

Many people also expressed their \textit{Anger, Anxiety, and Disappointment of Violence Against Women}. Unlike other social security events, the Tangshan incident was perceived as against women partially because of its sexual-harassment nature. Since some women users had already become wary of men, this event made their worries real. It struck Chinese women's nerve not just because the victims were women but also because the place, a barbecue restaurant, was too common in China, implying such events could be universal and happen in everyday encounters. 

Besides expressing fear and anxiety, some users accused the suspects of shockingly disrespecting and insulting the victims. They were angry because the man suspects regarded the women as vents for libido. The discussion also generalized to denouncement towards other men who disrespect women or even all men. They criticized men, especially those inclined to objectify women or show no empathy or understanding of women's disadvantaged reality.

The users also shared \textit{Anecdotes, Similar Cases and Warnings} with other women. One user mentioned that the incident reminded them to share an experience of being sexual-harassed that would otherwise have ``\textit{rotten in the stomach.}'' The disclosed experiences described personal stories and often were accompanied by warnings to other women and an appeal for social support within woman groups. They warned others that such violence could befall any single woman and is not predictable: \textit{``She is more than just her, she is us.'' ``If we don't speak up this time, we may not have the opportunity to speak up next time.''} These posts gained noticeable popularity. In general, they leveraged such empathetic narratives to stimulate women's sense of their reality and, simultaneously, attempted to build social support among those sharing similar experiences.

\subsubsection{Irrelevant or Ineffective Content} 
Although this content class constituted 67.42\% of the non-empty posts, most were not information-rich. Topic modeling clustered posts in the \textit{Irrelevant or Ineffective} class into ten topics, among which the major ones include Sentimental Expression of Frustration, Asking for Follow-up and Attention, Curses to the Suspects, Questioning the Ability of Local Police, and Irrelevant Content (not even related cases). Fully reporting this content class would make the analysis deviant from the focus of this study, so we only highlight the topic of \textbf{Sentimental Expression of Frustration}. This topic consisted of 22.81\% posts in this class with the following keywords: Tangshan, society, incident, hope, beating, event, people, hot search, nation, law, severe punishment, justice, hotness, China, disappointment, world, happen, attention, victims, law enforcement. Examples are given below:
\begin{quote}
``\textit{Seeing everyone keep posting about the beatings, hoping for an explanation, is like seeing the Little Red Mansion that filled the screen, but in the end there was no news and it was like nothing happened... It was covered up by trending searches of irrelevant entertainment and disappeared in the Internet wave like countless incidents that caused public outrage...}''

``\textit{So will our persistence be useful? Will it ever be okay?... it won’t take a few days for the Tangshan incident to be forgotten... After the incident, I found that it did not serve as a warning, and the incidents alike that followed continued unabated.}''

\end{quote}
Typically, these posts manifested the commonly-held frustration and exhaustion among the participants (including grassroots feminists), often followed by distrust of enforcement agencies. The participants felt that shortly the discussion would vanish, their impact was trivial, and how little progress or change they made after times of similar hysteria.

\subsection{RQ2: Discursive Patterns Around Gender Dimension}
The legitimacy and importance of gender factors were in the core conflict between the criminal and gender perspectives. Those \textit{Crime-related} posts tended to deny the incident's relation to gender. In contrast, \textit{Gender-related} posts necessitated generally feminist interpretation. There were also attempts to mediate the conflicting sides, making the point that criminal and gender interpretations are not necessarily mutually exclusive. In this subsection, we report their discursive patterns by the following themes: (1) Objection to Gender Dimension, (2) Support to Gender Dimension, and (3) Bridging the Two. In the heading for each pattern, we included the proportion of posts falling into that pattern or overarching theme. The proportions for patterns are relative to the overarching themes, and the proportions for the themes are relative to all the sampled posts. A post might fall into multiple patterns. In addition to what was reported below, we also found common discourses of stigmatization against feminism explored by previous studies \cite{mao_feminist_2020}.

\subsubsection{Objection to Gender Dimension (43.0\%)}
The vast majority of posts showing a hostile standpoint against advocates of gender factors were from the \textit{Crime-related} class. Those posts were inclined to ignore, deny, or downplay gender factors in interpreting the incident. Various discursive patterns emerged against feminist opinions, often intertwined and combined. The objection usually converged with legitimized criminal interpretation or re-framed the incident's conflict to those other than gender. The posts also criticized the feminists as being ``\textit{extreme}'' and ``\textit{inciting conflict between the genders}'' or ``\textit{inciting gender opposition}.'' Following are the major discursive patterns used to object to the gender dimension.

\paragraph{\textbf{Re-framing Conflict} (27.1\%)} 
A criminal perspective prevailed as the incident involved illegal gangs and violence against the law. However, often the incident was re-framed as a purely criminal issue, which denied gender factors and discussions. Even though gender factors were not completely disavowed, some would confer a priority---criminal topics precede gender topics. They usually re-framed the incident's nature to the conflict between the ``\textit{law-abiding and law-breaking},'' and thus defined the incident as an ordinary social security event. 
Some posts sought endorsement from ``\textit{equality before the law}'' by referring to that judicial practice about violence does not distinguish an abuser's gender, and thus gender factors were out of the discussion. Such an argument only stressed equality but missed equity and the fact that modern laws have provisions specifically for protecting women.

Besides legal framing, some related their opinions to normative social ethics about evil and good. They typically claimed that ``\textit{No discussion of men and women, only good and evil},'' where the evil was generalized from the gangsters, and the good was from the victims. Such discourse deducted and regressed the debate since, as some other posts pointed out, the intent was to cease gender-related and feminist expressions in the name of commonly-held ethics.
Some posts also related to the conflict between strong and weak: ``\textit{The protection is for the vulnerable, not just women; the punishment is for the evil, not men.}'' Again, the strong/weak pattern ignored gender factors and particularities associated with gender. Those posts were similar in that they attempted to fit the incident into a general and conservative value framework while refusing gendered reality. One example is as follows:
\textit{
\begin{quote}
``...
There still are other people getting hurt: may be boys, the elderly or even children. So it's not only women who will be hurt, but that the four women victims belong to a vulnerable group in front of these abusers.''
\end{quote}}

Women in this post were assembled with other groups into the umbrella term of ``\textit{the vulnerable}.'' Thus the particular power relations between genders in a patriarchal society were dismissed, and the frame shifted. Also, the post used the possibility of violence toward other groups as an argument, making the incident seem indifferent to an ordinary violent event. The substitution-like discursive pattern is further explained in \textit{Equalizing Vulnerability Between Genders}.

\paragraph{\textbf{Equalizing Vulnerability Between Genders} (10.2\%)}
This type of discourse could be characterized as ``\textit{violence does not discriminate between genders or among other social groups}.'' In other words, violence could equally befall each person regardless of their gender and intersectionality, depicting gender as an insignificant factor. For example, ``\textit{Even if for men, it's possible to be beaten for looking at each other ...}'' These posts shared an underlying assumption that women encounter violence for the same reason and mechanism as men do.

\paragraph{\textbf{Uniting All: ``people help people''} (4.9\%)}
An English phrase ``\textit{people help people}'' appeared in some posts to replace ``\textit{girls help girls},'' which was generally perceived as feminist. The posts envisioned a utopia where everyone unites and provides mutual help in facing any crisis, as one post mentioned:
\textit{\begin{quote}
``I hope that in the face of violence, in the face of any other injustice, men and women can stand in the same position and angle to help the disadvantaged. I hope not just `girls help girls,' or `boys help boys,' but `people help people.'''
\end{quote}}
Although appealing on the surface, such a narrative did not become aware of gendered lived experiences and systematic inequality.
The phrase ``\textit{people help people},'' in this particular context of gender-related violence and sexual harassment, was misleading because it suggested ignorance of neither contingent nor systematic gendered reality, and it would not be achieved with gender issues set aside. In this way, the seemingly prospective vision might only achieve its reverse.

\subsubsection{Support to Gender Dimension (54.24\%)}
All posts supporting gender-related and feminist discussion belong to the \textit{Gender-related} class. The posters adopted varied discursive patterns to defend the necessity of gender factors in the incident. A common characteristic among them was their awareness of the dissimilarity between genders in a social and cultural sense.

\paragraph{\textbf{Unmasking Sexual Motive and Challenged Masculinity} (33.1\%)} Based on the video footage, people argued that the oral conflict and the subsequent beating originated from the man's sexual harassment, but the motive was less mentioned among the crime-related posts. Thus, these posts related the incident to gender by assuming that women are victims in scenarios of sexual harassment. A post implied that criticizing sexual motive is the reason behind gender discussion: ``\textit{Isn't the reason why this happened because of sexual harassment, the problems that arise between the genders?}'' Another poster advocated that gender factors were precedent to criminal factors: 
\textit{\begin{quote}
``We CAN talk about human rights, public security, and the social system, but in this case, the root cause is sexual harassment. The physical identities of men and women are satisfied first, and then the social identities of abusers and victims are satisfied.''
\end{quote}}
Here, the sexual motive was regarded as the ``\textit{root cause}'' that was based on the ``\textit{physical identities of men and women}'' or their relations. Violence toward women due to sexual motives and gender preconception is clearly a target for feminist criticism because of its objectification of women and ignorance of women's will. Gender factors were defended by linking sexual motives with broader and structural gender issues.

Moreover, for some posters, the women victims' resistance challenged masculinity so that the violence occurred to maintain domination over the women and ``\textit{face in front of other men.}'' The conflict originated from the abuser's sexual harassment, which was resisted by the woman victim asking, ``What are you doing?'' The man replied with ``I wanna f*** you.'' Then the woman's friends hit the man, which triggered the man's outrage and violence.
Based on observable details, some posters thought the man's violence did not target who had beaten him but who was inferior to him. When challenged, the man's reaction was not to fight back but to maintain his nominal superiority. They also pointed out that the man's reaction might be motivated by his anxiety of losing face in front of others, especially his man friends. Since the man's friends had been informed of his intention and others around noticed the quarrel, the man might find himself awkward when the woman refused, and only continuing to demonstrate masculinity and power could conceal his failure. Some pushed the analysis deeper by saying that the suspect who failed to seek privilege over women feared being regarded as a loser and being expelled from the male privilege group.

\paragraph{\textbf{Redressing Biased Media Rhetoric} (30.3\%)}
The media's report did not mention sexual harassment but used other phrases for replacement. In the mainstream and smaller media news reports, ``\textit{harassment}'' was relatively rare, and ``\textit{sexual harassment}'' was even less. Mainly used phrases include ``\textit{picking up a conversation},'' ``\textit{chat},'' and ``\textit{stroke the back},'' all are less severe than ``\textit{sexual harassment}.''
\begin{quote}
``\textit{This is what the official media want to cover up, so they replace the word `sexual harassment' with neutral words such as `stroking the back' and `picking up a conversation.' The official media's comment on the incident is the key reason why women continue to be angry. The institutions of public authority erased this obvious gender element, isolating it as a contingency... 
}''
\end{quote}
As exemplified by this post, many expressed anger and dissatisfaction with media rhetoric that downplayed the nature of sexual assault. Criticisms against the domestic media thus constituted one of the facets of feminist advocacy in the debate.

\paragraph{\textbf{Refuting Ignorance of Sexism} (9.7\%)}
Proponents of the gender perspective also emphasized that the incident was a realistic representation of men's ignorance of sexism. Some posts even made an analogy between gender blindness and color blindness, drawing their connection to argue for the necessity of involving gender perspective:
\begin{quote}
\textit{``The man's argument that this incident has nothing to do with gender is equivalent to the White American policeman knelt down on the black man to death, but saying it was just excessive police enforcement and had nothing to do with skin color.}''
\end{quote}

This post provided narratives in which race was assumed to take a similar role as gender in the Tangshan incident---both were situated in the confrontation with criminal frames. The post implied that if one admits that race took effect in the scenario described, one should also consider gender factors in the Tangshan incident. One post pointed out that the Chinese foreign ministry spokesperson mentioned that the murder of George Floyd ``\textit{reflected the severity of racial discrimination and violent police enforcement in the US;}'' however, the Chinese government did not explicitly mention support or refutation of gender factors in the Tangshan incident.

\paragraph{\textbf{Challenging Equal Vulnerability Between Genders} (8.5\%)}
Some posters refused the discourse that men have equal chances and reasons for being attacked by others as women do. They stated that men are less likely to experience sexual harassment and violence from women as a way to refute the idea of equal vulnerability between genders. These discourses reflected the grassroots feminists' understanding of the qualitative differences between how violence against women and men occurs.
\begin{quote}
\textit{``In normal situations, does a girl threaten the life of a boy?...}
But the fact is that normally, boys won't get into that situation, boys who want to resist can fully protect themselves.''
\end{quote}
They thought that ``\textit{men do not get into that situation ... men can protect themselves},'' and women would be more passive, which is not due to the biological determinants but rather the social context where men's actions receive more endorsement and toleration. Men, due to limited experience, are often unaware of their privilege and the subordinate conditions experienced by women at both individual and social levels. Men most likely ignored the hierarchical gender relations so they would think gender equality was well achieved, and thus attributed women's encounters to their individual ``\textit{misbehavior}'' or ``\textit{bad luck}.''
\begin{quote}
\textit{``If they don't resist, men may be beaten to death, but what else will women encounter besides being beaten to death? The cause of the incident has already determined the risk and degree of persecution of women}.''
\end{quote}
Some posters also mentioned that women are regarded as the vent for men's libido in addition to brutality, which profoundly shaped their inferior position relative to men.

\subsubsection{Bridging the Two (11.9\%)}
Besides the dichotomized parties of objection and support to gender-related discussion, another group acknowledged the criminal aspect while being aware of gender issues. Most posts belonged to the \textit{Crime-related} class. Their posters thought that criminal and gender aspects were not exclusive. In principle, both problems can be solved without conflict---attention to criminal issues and legal institutions is necessary to control crime, but people should not deny gender-related or feminist interpretations. One post of this kind goes as follows:
\begin{quote}
``\textit{The nature of the vicious wounding incident is not `hatred to women', but `gangs oppressing people', which is common and not so closely related to gender. 
However, this is not to say that gender is not related to this incident - the weak social status of women and the contempt for women that have permeated the culture are one of the many reasons, and it is also universal and cannot be ignored by the criminal nature one bit! ...
Sweeping away the soil and protective umbrella that breeds underworld forces from the system, and ideologically sweeping away the misconception that bullying the weak (women) are both indispensable (the latter is also the same when it is replaced by migrant workers or the elderly).}''
\end{quote}
For them, both gangs and harassment were the causes of the violent incident. Therefore, some proposed eliminating the illegal gangs and umbrellas in the government and admitted the necessity of ``\textit{ideologically sweeping away the misconception}'' toward women.

\section{Discussion}

Sexism has shifted from overt dehumanization to covert and subtle forms such as gender blindness \cite{stoll2013race}. With the development of civil society, overt sexist language and sentiments are less expressed to adhere to social taboos. However, the conservative cultural norms around gendered roles, representation, and all the values coming along are still ingrained. In \ref{sec6_1}, we first add to the critique of gender-blind sexism reproduced on social media by exposing and discussing how the new sexist ideology moralizes itself by connecting it to normative moral convictions. We then discuss in \ref{sec6_2} a range of restrictions for conducting digital feminist activism on Weibo and Chinese social media in general. Based on the above discussion, we propose implications for design and research around online activism in \ref{sec6_3}.

\subsection{Normalization of Gender-Blind Sexism}
\label{sec6_1}
According to Stoll's initial framework, gender-blind sexism consists of four major frames: abstract liberalism, cultural sexism, minimization of sexism, and naturalization \cite{stoll2013race}. In the case of the Tangshan incident, which is a mix of criminal and gender issues, the majority of sexist discourses centered around the minimization of sexism (downplaying the role of sexism in determining the life chances of women \cite{anskat_dissemination_2021}). Abstract liberalism (stressing equal opportunity and individual choice among genders), cultural sexism (attributing gendered differences to culturally based arguments), and naturalization (naturalizing gender stratification through sociocultural and biological ``evidence'') appeared rarely. One explanation is that the discussions focused more on interpreting the specific case where confrontations between genders in a broader and deeper sense, on which other frames of gender-blind sexism base, were not quite visible. Instead, we found that people frequently used normative (re-)framing as a persuasive technique to minimize sexist attributes. Adherents to the frames positioned themselves at a moral highland backed by common values. The findings, therefore, further contextualize Stoll's framework of gender-blind sexism in a real-world case and demonstrate how such sexism is entangled with normative moral convictions.

Moral framing, according to Lakoff \cite{Lakoff1996_LAKMPW}, refers to the framing of which the driven force is based on moral concerns. Research across disciplines about moral framing suggests it is an effective persuasion tool, especially in political debate \cite{feinberg_moral_2019, voelkel_morally_2018}. The closer it is between the framing and the persuasion target's morality, the more likely the persuasion will succeed \cite{feinberg_moral_2013, feinberg_gulf_2015}. As illustrated in our findings, normative moral convictions supporting social security, vulnerable groups, and good citizens were adopted to affirm the legitimacy of dismissing feminist discourses and women's particular concerns. This was evident in the idea of ``\textit{people help people},'' which paradoxically made normally unacceptable opinions (e.g., people should not pay \textit{special} attention to women being harmed by men) seemly reasonable. These common moral values were, in essence, not contradictory to feminism, but through such rhetoric, they endorsed the minimization of sexism. This phenomenon also affirmed the necessity of studying how discursive techniques obscure and sugarcoat gender-blind sexism. Besides, Kodapanakkal et al. \cite{kodapanakkal_moral_2022} supported the statement that moral frames and re-frames also decrease the willingness to compromise, suggesting that the underlying gender-blind sexism may be more persistent when pivoted on normative values.

\subsection{Multifaceted Restrictions of Feminist Activism on Weibo}
\label{sec6_2}
Social platforms mediate the production and consumption of user-generated content, and naturally their features influence how ideologies disseminate and reproduce online \cite{anskat_dissemination_2021}. Matamoros-Fernández \cite{matamoros-fernandez_platformed_2017} proposes the concept of \textit{platformed racism} as the racism generated from platforms' designs, affordances, policies, and culture of use. Wu et al. \cite{wu_conversations_2022} further examine how the designs, policies, and content of Reddit support racism, resulting in weaponized identities and digital gentrification. They also suggest contextualized interventions and governance. When it comes to gender issues, Adrienne \cite{massanari_gamergate_2017} investigates how Reddit's karma point system and other mechanisms support anti-feminist and misogynistic activism. In the above work, dominant ideologies utilize the functionalities of platforms. Informed by this prior work, in this subsection, we contribute to the discussion by offering a brief \textit{interaction critique} \cite{Bardzell2008Criticism} where we highlight the restrictions of conducting digital feminist activism on Weibo, including platform values and cultural symptoms. We believe understanding the complexity of the broad context is inevitable for any meaningful and practical design intervention to be proposed. Although focusing on a single platform, the insights can be transferred to analyzing other cases in other contexts sharing similarities.

\subsubsection{Neoliberal and Political Rationality of Online Platforms}
It is critical to highlight how Weibo's rationale, embedded values, and culture affect the ideological manifestation. They reveal the foundation of (re-)producing gender-blind sexism on the platform. Informed by literature depicting the commodification and commercialization of Weibo within the shifting cultural politics of China towards regulation of discontent \cite{jia_tracing_2020, benney2017decline}, we argue that Weibo is primarily driven by a neoliberal and political rationality, which represses the visibility of feminism and the platform's potential as a feminist site.

Weibo, a commercial social computing platform, positions financial benefits as the highest priority and one of the ultimate rationales for all decisions. Weibo profits mostly from selling marketing channels, subscription services, and financial products, which all depend on user volume and engagement. The goal is to attract users' attention as elaborated by the idea of attention economy where human attention, as a ``focused mental engagement on a particular item of information'' is treated as a limited resource and currency \cite{davenport_attention_2001}. Likewise, users' engagement functions as currency on platforms like Instagram within its ``neoliberal visual economy'' \cite{mahoney_is_2022}. Weibo has deployed multiple techniques to achieve this goal. \textit{Weibo Hot Search}, a feature of Weibo consisting of real-time frequently searched keywords, manifests timely attention at the surface but actually incentivizes frequent, ephemeral production of and competition among top-selling celebrity gossips, sentiments, and opinions \cite{jia_tracing_2020}. When users write posts, hashtags are listed and recommended in an order calculated by an algorithm. They may tend to use top-ranked or recommended hashtags, making the popular more visible. On the contrary, other hashtags would receive significantly less attention. Such designs are for congregating public attention and increasing revenue. Content value evaluation is based on quantified user volume, time spent, shares, comments, likes, and others. This leads to pervasive cultural and ideological unawareness of Weibo as an information producer and distributor. At its root, Weibo creates a neoliberal venue where information posted by different individuals competes for user attention and engagement. Within such neoliberal logic, hierarchies of information and opinions are natural and taken for granted; domination and subordination are due to objective reasons like truthiness and quality. Thoughts as unpopular as feminism on Weibo therefore face critical barriers to entering the central discursive realm because the mainstream ideologies integrated with the neoliberal economy of Weibo, restricting feminist rhetoric from dissemination. Interestingly, as feminism spreads among the public and enters popular culture, enacting the era of ``post-feminism,'' the ``feminist'' content gaining the most attention is that posing little challenges to normative social structures of heterosexuality, patriarchy, race, and class \cite{banet-weiser_empowered_2018}. It is not hard to recognize the conservative power of the neoliberal economy exerted on feminist expressions on Weibo.

Nevertheless, Weibo certainly cannot make decisions exclusively based on economic calculation but ignores the political context where it is situated. The two rationales are constantly in tension and reflect the broader conflict between the free market and authoritarian domestic politics.

Chinese internet has been known for enforcing content censorship and firewall blocking some overseas IP locations, and Weibo is not an exception \cite{bamman_censorship_2012}. Weibo needs to monitor user-generated content and trends of public discussions, execute censorship, block accounts when it identifies the need to control,  reduce the risk of turmoil, and hide ``unpleasing'' reports about or criticism towards local or central authorities \cite{zeng2017social}. Chinese authorities pay special attention to political dissidents and ideologies from Western nations, including feminism. Largely, such action is due to political pressure from the authorities on Weibo, which possesses the power to shape public opinions. However, this is not to say feminism is wholly blocked or incompatible with the current Chinese political realm; in fact, China has a history of supporting women's rights and emancipation \cite{chen2011many}. The problem is the authorities' concern with the potentially subversive power of feminism, avoiding gender topics in the public realm but accumulating tensions with the growing popularity of feminism and awareness of women's issues. These tensions are manifested in the discourses on Weibo, pushing against censorship and anti-feminism.

\subsubsection{Official and Public Sensitivity to Feminism}
The low visibility of feminist discourses may be one of the major caveats under Weibo's current designs and policies. However, Chinese feminists have been striving to extend their reach on social media and establish alliances to inform broader audiences \cite{wang_chinese_2019} under not only pressures from authorities but also cultural stigmas from the public who repress feminist agendas on Weibo.

The semantic mismatch between the hashtags and the text content suggests a status of feminism and feminist identity on Weibo as unspeakable. Hashtags similar to \#\textit{girls help girls} did not gain popularity, whereas posts using neutral hashtags such as \#\textit{Tangshan beating} but with text supporting ``\textit{girls help girls}'' appeared frequently. Hashtags with an explicit feminist stand or identity were not used much by the grassroots feminists, but their ideas were expressed through posting with neutral or ``politically correct'' (meaning adherent to mainstream framing to see the incident as purely a criminal issue) hashtags to engage in discussion. Although there were hashtags with gender awareness like \#\textit{Violent attack of Tangshan girls is a nightmare for all}, the discussions are much less than others. Compared to \#Black Lives Matter and \#MeToo, the most popular hashtags where identities are made explicit and speakable (Black, Me), whereas \#\textit{Tangshan beating} is impersonal and embeds a governmental perspective. It is difficult for us to imagine the most popular hashtag being one with a feminist statement or identity, given the authorities and the public's sensitivity.

A possible explanation for this could be the policy enforcement by Weibo and the governmental bureau. The hashtags with an apparent feminist stand could be restricted or censored to control public opinions. It could also be due to the intention of Weibo users to avoid publishing ``aggressive'' posts amidst the sensitive Chinese Internet or to avoid accusations of being ``pastoral/rural feminists'' (a form of stigmatization to feminists in China, accusing feminists as only seeking privileges but not performing duties. Such ``feminists'' exist in China as well do false accusations) \cite{mao_feminist_2020}. The ultimate consequence of both official and public pressures was the unspeakable status of feminism on Weibo. With that said, feminism being unspeakable does not mean it cannot spread on Weibo; rather, as illustrated in our findings, women were consciously utilizing feminist discourses to argue and affect others. Therefore, instead of promoting feminist hashtags and identities explicitly, feminism on Weibo is growing beneath the surface as an undercurrent. But going back to Cole's argument that hashtags can be liberating \cite{cole_its_2015}, we argue that the absence of concentrated feminist rhetoric (e.g., hashtags with an explicit feminist stand) would be a limitation for the current digital feminist endeavor in China.

\subsubsection{Unrepresentative Online Participation}
This digital feminist activism still poses the issue of representation in terms of whose voice got raised and heard and whose interests were taken care of. User demographics and discourses were urban-centered in the online discussions of the Tangshan incident. In our data, economically developed provinces/cities like Beijing (0.12\% population posted about the incident on Weibo) and Shanghai (0.08\%) tended to have higher participation ratios than less-developed areas like Qinghai (0.02\%) and Yunnan (0.02\%) during the discussion, based on population data from National Bureau of Statistics of China \cite{nationalbureau}. This aligns with the result from Weibo's user development report that Weibo is used mainly by users from economically developed regions of mainland China, born in the 90s and 00s, holding a higher education degree and a white-collar job \cite{WeiboUser2020, WeiboUser2016}. Most participants shaped the focus of discourse toward urban spaces, workplaces, and schools, mostly in the public realm.

What was missing from the present discourse of digital feminist activism on Weibo were voices from women who could not participate in the discussion due to various barriers. Lack of Internet access and technological literacy, for example, can be crucial barriers for rural women in less-developed regions of China to raise their voices online. Such engagement also requires ample leisure time devoted to the Internet to browse, search, read, respond, discuss, and sometimes argue with others back and forth, but apparently, this leisure is a privilege belonging to the leisure class exempted from economically productive work. Although experiencing restrictions rooted in the same patriarchal structures, it may still be difficult for urban, young, and highly educated digital feminists to imagine the lived experience of those excluded from online expressions yet facing even harsher conditions.

This relates to the growing discourses within the HCI/CSCW community regarding intersectionality ``as a framework for engaging with the complexity of users' and authors' identities, and situating these identities in relation to their contextual surroundings'' \cite{schlesinger2017intersectional}. Awareness of intersectionality and overlapping identity categories is seen in previous HCI/CSCW work, such as studying homeless mothers' use of deployed computing system \cite{Christopher_Part_2012} and e-commerce live-streaming practices of Chinese rural women \cite{Tang_Dare_2022}. By reviewing some of these works, Schlesinger et al. highlighted the importance of reporting the context of the study and demographics in detail to understand the situations facing the participants/users holistically \cite{schlesinger2017intersectional}. Although they mostly speak to the research community, this has implications for activists as well. Our concern about representation strongly resonates with what Moitra et al. found about how participating in the \#MeToo movement in India embedded existing social divides and power relations of caste, class, gender, and sexual orientation, which impeded some women from participating because of the inability to align with the elite participants \cite{moitra2021parsing}. It is therefore imperative for digital feminist activists to reflect on their positions and perspectives and be aware of issues of intersectionality and representation in digital activism in order to promote equity and justice for all women with various distinct social identities within intersecting contexts.

\subsubsection{Ephemerality and Frustration of Participation}
The posts expressing uselessness and frustration of participation characterized the common mental state of engaging in online activism for social change. These expressions suggested a potential platform-wise cause of such sentiment---the ephemerality of online information and social discussion. On the one hand, participants wished for the discussion and public attention to be sustained; on the other, they wished for practical effect. But reality was against their wishes, and the same situations repeated over and over again, rendering their perceived inability to change. This differs from literature claiming self-gratification of individuals after participating in politically ineffective online activism, or slacktivism \cite{christensen2011political} and contributes to HCI/CSCW as a new side effect of slacktivism \cite{lee2013does, rotman2011from}. Apparently the participants in the discussion of the Tangshan beating incident were not satisfied with the outcome of their effort. Some forms of feedback and visible outcomes are critical for grassroots feminist activists to continue their endeavors amidst the restrictions. Possible forms could be increased visibility of feminists and women's rights organizations, expanded rational discussions on women's concerns, official responses concerning women's particular circumstances, and acknowledgment of the importance of feminism. This substantial progress may encourage grassroots feminists to move forward.

\subsection{Implications for Design and Research}
\label{sec6_3}
Reflecting on our results and the literature, we provide the following design and research implications for the CSCW community.

\subsubsection{Enable Peer Support Among Grassroots Feminists}
Social platforms should support vulnerable and disadvantaged groups by protecting their voices from marginalization again in online spaces. Grassroots feminists, in the case of the Tangshan incident, were trying to establish connections with others as a way of resistance. Although the network was fragile and not organized like some other online activism \cite{tufekci_twitter_2017}, the grassroots feminists were sharing anecdotes and sympathy and constructing solidarity rhetoric to resist gender-blind sexism and combat traumatic experiences. Besides, frustration after participating in politically ineffective online activism also implies a design need for dealing with such emotional and mental challenges. Based on the observations and following the principles of trauma-informed design \cite{chen2022trauma, scott2023trauma}, we propose that establishing peer support is the intrinsic need for Chinese grassroots feminists to navigate the restricted online environment in China sustainably. Through peer support, the participants can establish mutual help through ``sharing stories, disclosing recovery journeys, and linking and normalizing their shared experiences'' \cite{scott2023trauma}.

\subsubsection{Hashtags Can be Less Robust for Research in Non-Western Contexts} 
Hashtags do not accurately represent opinions, as seen in the mismatch between hashtags' and content's semantics. Not only because a hashtag is too short to convey the full meaning of the post body, but more importantly, that hashtag is more for alouding the words than being the words themselves, and that alouding is twisted by the cultural politics it embeds in.

One of the most common reasons for using hashtags in online activism is its ability to get public attention, which is often further elaborated as the liberating potential for marginal communities \cite{cole_its_2015, losh2019hashtag, simpson2018integrated}. This prevalent positive impression on the potential of hashtags is from and for Western contexts and needs caution when being transferred to analyzing hashtag use in non-Western contexts. The hashtag \#MeToo failed to encourage participation from Indian women who could not align with the embedded ``Western values'' \cite{moitra2021parsing}. Therefore, strictly following the research procedures done in Western contexts and drawing conclusions simply from hashtag use would miss the experience of the non-participants. The public and Chinese authorities will likely attack in China, explicit and highly visible feminist identities and expressions. Feminist hashtags can attract public attention to feminism but also regulation and enforcement from the authorities, rendering the feminists disadvantaged. On the contrary, using neutral-toned yet popular hashtags to publish feminist or gender-related content can be a tool to lower the feminist profile to avoid cyberbully and censorship amidst sensitive cultural politics. Researching hashtags in non-Western contexts warrants contextual understanding beyond posting behavior and platform features to include the variety of social dynamics -- something ``external'' to individual interaction. Concluding based merely on hashtag use in such contexts would fail to achieve robustness in research. Taking our experience as an example, when conducting initial analyses, we tried to visualize the relative positions among crime and gender-related hashtags in a co-occurrence network of hashtags. The gender-related hashtags were far away from the network center and had small numbers of posts, potentially leading to the wrong conclusion that gender-related expressions were significantly missing in the online discourse. Only after analyzing the text content posted were we able to observe a comparable volume of crime and gender-related discussions and the mismatch between hashtag and content semantics.

\section{Limitations and Future Work}
This work's limitation unfolds into two parts. First, we could only scrap a subset of all posts due to the limitation set by Weibo (e.g., displaying only 1,000 posts for each hour when searching by hashtags, but more posts got published during that hour, and we could not precisely measure the loss). And we could not collect posts that did not use hashtags or did not contain the keyword ``Tangshan'' in the hashtags but still talked about the incident. Second, there was censorship by Chinese authorities. Posts on Weibo might be censored based on content sensitivity, language, and word usage, IP location, and user records. Tracing content censorship, we found posts blocked from access via desktop websites and previously visible posts later marked as ``deleted by the author'' and becoming invisible.

Based on the findings and discussion, future work could be: (1) analyzing the social collaboration network of the women users to examine the dissemination mechanism, relation, and influence dynamics; (2) interviewing people who participated in digital feminist activism about their experiences and perceptions of gender-blind sexism and restrictions of conducting feminist activism online; and (3) investigating hashtag use in contexts outside of Western nations.

\section{Conclusion}

By employing computational methods and feminist critical discourse analysis on the online debate of a gender-related violence incident in China, we uncovered the trends, topics, and discursive patterns framing the incident's criminal and gender dimensions. 
The mixed method approach captured the collective patterns while maintaining a nuanced understanding of meanings emerging in online discourses. Results from computational methods provided insights into the large-scale content divisions and facilitated the following qualitative analysis. The qualitative method brought a feminist perspective to uncover implicit meanings and values.
We found that gender-related and feminist discourses were suppressed by criminal frames in the discussion to minimize sexism, and Chinese grassroots feminists were actively engaging in the discussion to push back and raise their voices. 
The criminal framing of the incident embedded the idea of gender-blind sexism, which continues to penetrate the digital spaces. We also highlight challenges in conducting digital feminist activism on social platforms in non-Western contexts like China and propose the implications for design and research.

\bibliographystyle{ACM-Reference-Format}
\bibliography{main}

\end{document}